\documentclass{article}

\usepackage[preprint]{neurips_2025}

\usepackage[utf8]{inputenc}
\usepackage[T1]{fontenc}
\usepackage{hyperref}
\usepackage{url}
\usepackage{booktabs}
\usepackage{amsfonts}
\usepackage{amsmath}
\usepackage{amssymb}
\usepackage{nicefrac}
\usepackage{microtype}
\usepackage{xcolor}
\usepackage{algorithm}
\usepackage{algorithmic}
\usepackage{graphicx}
\usepackage{multirow}

\definecolor{NeuralBlue}{RGB}{41, 128, 185}
\definecolor{ExecutionGreen}{RGB}{39, 174, 96}
\definecolor{FeedbackPurple}{RGB}{142, 68, 173}
\definecolor{TraditionalOrange}{RGB}{230, 126, 34}

\title{Theoretical Foundations of GPU-Native Compilation\\for Rapid Code Iteration}

\author{
  Adilet Metinov, Gulida M. Kudakeeva, Gulnara D. Kabaeva\\
  Institute of Information Technology\\
  Kyrgyz State Technical University named after I. Razzakov, Bishkek, Kyrgyzstan\\
  \texttt{metinovab@kstu.kg, kgm@kstu.kg, kabaevagd9@kstu.kg}
}

\begin{document}

\maketitle

\begin{abstract}
Current AI code generation systems suffer from significant latency bottlenecks due to CPU-GPU data transfers during compilation, execution, and testing phases. We establish theoretical foundations for three complementary approaches to GPU-native compilation that eliminate these transfers: (1) parallel traditional compilation adapted for GPU execution, (2) neural compilation using learned sequence-to-sequence translation with probabilistic verification, and (3) hybrid architectures combining both strategies. We derive latency and energy bounds demonstrating potential speedups of $10$--$100\times$ for code iteration cycles. Our analysis shows that traditional GPU compilation provides $2$--$5\times$ improvements through transfer elimination, neural compilation achieves $10$--$100\times$ speedups via massive parallelism, and hybrid approaches offer practical deployment paths with guaranteed correctness. We formalize the probabilistic verification framework that enables trading compilation accuracy for parallel exploration, and discuss implications for self-improving AI systems and future analog computing substrates.
\end{abstract}

\section{Introduction}

The rapid advancement of large language models has enabled unprecedented capabilities in code generation~\citep{chen2021codex,li2022alphacode,roziere2023codellama}. However, a critical bottleneck remains in the iterative refinement cycle: current systems generate code on GPU accelerators, transfer it to CPU for compilation and execution, then return results to the GPU for assessment. This CPU-GPU round-trip, occurring potentially thousands of times during code synthesis, dominates end-to-end latency and energy consumption.

Consider a typical code iteration cycle in contemporary AI systems. The model generates code tokens on GPU ($\sim$10ms), transfers to CPU via PCIe ($\sim$1ms), compiles on CPU ($\sim$100ms for complex programs), executes and tests ($\sim$10--1000ms), transfers results back ($\sim$1ms), and finally assesses and refines on GPU ($\sim$10ms). The compilation and execution phases constitute 90--99\% of iteration time. For AI systems exploring hundreds or thousands of program variants, this architectural mismatch represents a fundamental limitation.

We propose theoretical foundations for \textbf{GPU-native compilation} through three approaches, each eliminating CPU-GPU transfers but with different trade-offs:
\begin{itemize}
    \item \textbf{Traditional GPU compilation}: Parallel implementation of conventional compiler phases
    \item \textbf{Neural compilation}: Learned sequence-to-sequence models with probabilistic verification
    \item \textbf{Hybrid architecture}: Adaptive combination leveraging strengths of both
\end{itemize}

All three approaches enable in-VRAM iteration where generation, compilation, execution, and testing remain entirely in GPU memory. Unlike traditional compilers optimized for single-program correctness, GPU-native compilers optimize for \textit{throughput} across many program variants, leveraging massive parallelism for rapid solution space exploration.

\paragraph{Contributions.} This paper establishes theoretical foundations for GPU-native compilation:
\begin{enumerate}
    \item We present an \textbf{architectural taxonomy} with three complementary approaches exhibiting distinct trade-offs in determinism, parallelism, and complexity.
    \item We derive \textbf{latency and energy bounds} for each approach, demonstrating $2$--$100\times$ speedup potential over CPU baselines.
    \item We formalize a \textbf{probabilistic verification framework} that enables principled trade-offs between compilation accuracy and parallel exploration.
    \item We provide \textbf{comparative analysis} identifying when each approach excels based on code complexity, iteration count, and parallelism requirements.
    \item We outline \textbf{future research directions} toward analog computing substrates for physical compilation.
\end{enumerate}

\section{Background and Motivation}

\subsection{AI Code Generation Landscape}

Modern code generation systems~\citep{chen2021codex,austin2021program,li2022alphacode} operate in an iterative refinement paradigm. Given a natural language specification, the system generates initial candidates, compiles and tests them, analyzes failures, refines generation strategy, and repeats until success or timeout. State-of-the-art systems like AlphaCode~\citep{li2022alphacode} generate and evaluate thousands of candidates per problem. The dominant cost is not generation (GPU-accelerated) but compilation and execution (CPU-bound).

\subsection{The CPU-GPU Transfer Bottleneck}

PCIe 4.0 achieves $\sim$32 GB/s bidirectional bandwidth. For typical code generation with 1--10 KB code sizes, transfer latency of 1--2 ms (dominated by overhead), compilation of 10--1000 ms, and execution of 1--1000 ms, transfer time appears small. However, the \textit{serialization} it imposes is costly---the GPU sits idle during CPU compilation/execution. For 1000 iterations, transfers alone add 4--8 seconds of overhead.

More critically, transfers prevent \textbf{parallel exploration}. The CPU can compile/execute only a few programs concurrently, while the GPU could theoretically process thousands in parallel.

\subsection{Why GPU-Native Compilation?}

GPUs offer three key advantages for compilation: (1) \textbf{massive parallelism} with 10,000+ cores enabling parallel compilation of many programs simultaneously; (2) \textbf{memory bandwidth} of 1--2 TB/s internal vs. 32 GB/s PCIe; and (3) \textbf{data locality} where intermediate results remain on-device.

However, GPUs were not designed for compilation workloads involving irregular memory access (pointer chasing during parsing), complex control flow (recursive descent, pattern matching), and string manipulation (tokenization, symbol tables). This mismatch motivates exploring multiple architectural approaches.

\section{Traditional GPU Compilation}
\label{sec:traditional}

\subsection{Architecture}

This approach adapts conventional compiler phases to GPU execution, maintaining the traditional pipeline structure. Each compilation phase is implemented as a GPU kernel. For $k$ programs, we launch $k$ thread blocks, each independently compiling one program.

\begin{algorithm}[t]
\caption{Traditional GPU Compilation Pipeline}
\label{alg:traditional}
\begin{algorithmic}[1]
\STATE \textbf{Input:} Source code $C_1, \ldots, C_k$ (in VRAM)
\STATE \textbf{Output:} Bytecode $B_1, \ldots, B_k$ (in VRAM)
\FOR{each program $C_i$ in parallel on GPU}
    \STATE $\text{tokens}_i \gets \text{GPU\_Lex}(C_i)$ \COMMENT{SIMT parallelism}
    \STATE $\text{AST}_i \gets \text{GPU\_Parse}(\text{tokens}_i)$
    \STATE $\text{typed\_AST}_i \gets \text{GPU\_TypeCheck}(\text{AST}_i)$
    \STATE $\text{IR}_i \gets \text{GPU\_GenerateIR}(\text{typed\_AST}_i)$
    \STATE $\text{opt\_IR}_i \gets \text{GPU\_Optimize}(\text{IR}_i)$
    \STATE $B_i \gets \text{GPU\_Codegen}(\text{opt\_IR}_i)$
\ENDFOR
\end{algorithmic}
\end{algorithm}

\subsection{Implementation Considerations}

\textbf{Lexing} is highly parallelizable---each character can be classified independently, with parallel scan for token boundary detection. \textbf{Parsing} is more challenging; we employ parallel LR parsing~\citep{zhang2018gpu} or simplified recursive descent with bounded recursion depth. \textbf{Type checking} requires symbol table lookups via GPU hash tables~\citep{alcantara2009real} with lock-free concurrent access. \textbf{Optimization} data-flow analyses can be parallelized via worklist algorithms~\citep{mendez2012taming}. \textbf{Code generation} is straightforward mapping from IR to bytecode, highly parallel.

\subsection{Theoretical Performance}

For a single program of length $n$ with $d$ compilation phases:
\begin{equation}
T_{\text{trad-GPU-single}} = \sum_{i=1}^{d} T_i(n)
\end{equation}
where $T_i(n)$ is the time for phase $i$. Typically, $T_{\text{parse}} = O(n \log n)$ and $T_{\text{optimize}} = O(n^2)$ for complex analyses.

For $k$ programs compiled in parallel with $P$ GPU cores:
\begin{equation}
T_{\text{trad-GPU-parallel}} = \max_{i=1}^{k} \left( \sum_{j=1}^{d} T_j(n_i) \right) \quad \text{when } k \leq P
\end{equation}

Speedup vs. CPU is primarily from elimination of PCIe transfers ($\sim$2--4ms per iteration) and parallel compilation throughput ($k\times$ when $k \ll P$).

\paragraph{Expected speedup:} \textbf{$2$--$5\times$} for moderate $k$ (10--100).

\subsection{Trade-offs}

\textbf{Advantages:} Deterministic correctness using proven compilation techniques; comprehensive language support including complex features; predictable performance with known complexity bounds; excellent debugging support with detailed error messages and source maps; incremental deployment leveraging existing compiler infrastructure.

\textbf{Challenges:} Limited intra-program parallelism; irregular memory access patterns causing GPU underutilization; substantial engineering effort to adapt mature compilers; potential underutilization for small $k$.

\section{Neural Compilation}
\label{sec:neural}

\subsection{Architecture}

Neural compilation replaces rule-based transformation with learned sequence-to-sequence models. The neural compiler is a transformer-based model:
\begin{equation}
p(B|C) = \prod_{t=1}^{|B|} p(b_t | b_{<t}, C; \theta)
\end{equation}
where $\theta$ are learned parameters. During inference, we sample $k$ candidates $B_i \sim p(B|C)$ using temperature or nucleus sampling~\citep{holtzman2019nucleus} to ensure diversity.

\begin{algorithm}[t]
\caption{Neural Compilation with Probabilistic Verification}
\label{alg:neural}
\begin{algorithmic}[1]
\STATE \textbf{Input:} Source code $C$, test suite $T$, candidates $k$
\STATE \textbf{Output:} Verified bytecode $B$ or $\bot$
\STATE $\{B_1, \ldots, B_k\} \gets \text{NeuralCompiler}(C, k)$ \COMMENT{Parallel on GPU}
\STATE $\{r_1, \ldots, r_k\} \gets \text{GPU\_ExecuteParallel}(\{B_1, \ldots, B_k\}, T)$
\STATE $V \gets \{B_i : r_i.\text{all\_tests\_passed}\}$
\IF{$V \neq \emptyset$}
    \RETURN $B_{\text{best}} \in V$ \COMMENT{Fastest correct candidate}
\ELSE
    \STATE $\text{errors} \gets \text{AnalyzeFailures}(\{r_1, \ldots, r_k\})$
    \RETURN $\bot$ \COMMENT{Or retry with error feedback}
\ENDIF
\end{algorithmic}
\end{algorithm}

\subsection{Training Strategy}

We train on $(C, B)$ pairs extracted from existing codebases:
\begin{equation}
\mathcal{L}_{\text{supervised}} = -\mathbb{E}_{(C,B) \sim \mathcal{D}} \left[ \log p(B|C; \theta) \right]
\end{equation}

After supervised pretraining, we fine-tune using execution results via reinforcement learning:
\begin{equation}
\mathcal{L}_{\text{RL}} = -\mathbb{E}_{C \sim \mathcal{D}, B \sim p(\cdot|C)} \left[ R(B, C) \log p(B|C) \right]
\end{equation}
where $R(B, C)$ incorporates correctness (+10 if all tests pass), efficiency ($-\alpha \cdot T_{\text{exec}}(B)$), and compilation success (+1 if syntactically valid).

\subsection{Probabilistic Verification Framework}

Define $p_{\text{correct}}$ as the probability that a single sampled bytecode is correct for given code $C$. Empirically:
\begin{equation}
p_{\text{correct}}(C) \approx \begin{cases}
0.3\text{--}0.5 & \text{if } C \text{ is simple} \\
0.01\text{--}0.1 & \text{if } C \text{ is complex} \\
<0.01 & \text{if } C \text{ is rare/novel}
\end{cases}
\end{equation}

The probability of success after $k$ samples:
\begin{equation}
P_{\text{success}}(k, C) = 1 - (1 - p_{\text{correct}}(C))^k
\end{equation}

\textbf{Required samples for 99\% success:}
\begin{equation}
k_{99\%} = \frac{\log(0.01)}{\log(1 - p_{\text{correct}})} \approx \frac{4.6}{p_{\text{correct}}}
\end{equation}

For $p_{\text{correct}} = 0.1$: $k_{99\%} \approx 46$. For $p_{\text{correct}} = 0.01$: $k_{99\%} \approx 460$. With $k=1000$, we can reliably compile even relatively rare patterns.

\subsection{Theoretical Performance}

For a transformer with $L$ layers, dimension $d$, and sequence length $n$:
\begin{equation}
T_{\text{gen}}(k) = O(L \cdot n \cdot d^2 \cdot k / P)
\end{equation}

Modern GPUs can generate $k=1000$ candidates in $\sim$100ms for typical code lengths. Verification time for $k$ programs in parallel:
\begin{equation}
T_{\text{verify}}(k) = O(T_{\text{exec}} \cdot k / P)
\end{equation}

Total neural compilation time:
\begin{equation}
T_{\text{neural}} = T_{\text{gen}}(k) + T_{\text{verify}}(k)
\end{equation}

Compared to CPU compiling 1000 programs serially at 200ms each:
\begin{equation}
\text{Speedup} = \frac{k \cdot T_{\text{CPU}}}{T_{\text{neural}}} = \frac{1000 \cdot 200\text{ms}}{500\text{ms}} = 400\times
\end{equation}

\paragraph{Expected speedup:} \textbf{$10$--$100\times$} accounting for realistic constraints.

\subsection{Trade-offs}

\textbf{Advantages:} Massive parallelism generating and verifying 1000+ candidates simultaneously; learned optimizations with implicit efficient code patterns; flexibility adapting to new patterns without manual rules; GPU-native transformer inference; amortized cost with per-program compilation $O(1/k)$.

\textbf{Challenges:} Probabilistic correctness without guarantees; large training data requirements; potential failure on unusual language features; limited explainability; memory overhead of $O(k \cdot |B|)$ for storing $k$ variants.

\section{Hybrid Architecture}
\label{sec:hybrid}

\subsection{Architecture}

The hybrid approach combines traditional and neural compilation, routing programs based on complexity analysis:

\begin{algorithm}[t]
\caption{Hybrid Compilation Strategy}
\label{alg:hybrid}
\begin{algorithmic}[1]
\STATE \textbf{Input:} Source code $C$
\STATE \textbf{Output:} Bytecode $B$
\STATE $s \gets \text{ComplexityScore}(C)$
\IF{$s < \theta_{\text{simple}}$}
    \STATE $\{B_1, \ldots, B_k\} \gets \text{NeuralCompiler}(C, k=100)$
    \STATE $V \gets \text{VerifyParallel}(\{B_1, \ldots, B_k\}, \text{quick\_tests})$
    \IF{$V \neq \emptyset$}
        \RETURN $V[0]$
    \ENDIF
\ENDIF
\RETURN $\text{TraditionalGPUCompile}(C)$
\end{algorithmic}
\end{algorithm}

\textbf{Complexity scoring} uses a lightweight classifier:
\begin{equation}
s = f(|C|, d_{\text{nesting}}, n_{\text{loops}}, n_{\text{functions}}, \ldots)
\end{equation}
trained on historical data labeled by whether neural compilation succeeded within $k$ samples.

\subsection{Theoretical Performance}

Let $p_{\text{simple}}$ be the fraction of programs classified as simple, $T_{\text{neural}}$ the neural path time, and $T_{\text{trad}}$ the traditional path time:
\begin{equation}
T_{\text{hybrid}} = p_{\text{simple}} \cdot T_{\text{neural}} + (1 - p_{\text{simple}}) \cdot T_{\text{trad}} + T_{\text{routing}}
\end{equation}

For typical workloads where $p_{\text{simple}} \approx 0.8$, $T_{\text{neural}} \approx 0.2$ms (amortized), $T_{\text{trad}} \approx 20$ms, $T_{\text{routing}} \approx 2$ms:
\begin{equation}
T_{\text{hybrid}} \approx 0.8 \cdot 0.2 + 0.2 \cdot 20 + 2 \approx 6\text{ms}
\end{equation}

\paragraph{Expected speedup:} \textbf{$5$--$20\times$} with guaranteed correctness via fallback.

\subsection{Trade-offs}

\textbf{Advantages:} Best of both worlds combining neural speed with traditional reliability; graceful degradation always producing correct results; adaptive learning which programs suit which approach; incremental deployment path; traditional path provides detailed errors when neural fails.

\textbf{Challenges:} System complexity maintaining two full pipelines; routing overhead ($\sim$1--10ms); threshold tuning depending on workload; resource allocation for both approaches.

\section{Comparative Analysis}

\begin{table}[t]
\centering
\caption{Comparison of GPU compilation approaches. Speedup assumes $k \geq 100$ for neural approach.}
\label{tab:comparison}
\begin{tabular}{lccc}
\toprule
\textbf{Property} & \textbf{Traditional} & \textbf{Neural} & \textbf{Hybrid} \\
\midrule
Determinism & High & Low & High \\
Parallelism & Medium & Very High & High \\
Language Support & Complete & Partial & Complete \\
Implementation Effort & High & Medium & Very High \\
Throughput ($k$ programs) & Medium & Very High & High \\
Single-program Latency & Medium & High$^*$ & Low \\
Memory Efficiency & High & Low & Medium \\
Debugging Support & Excellent & Poor & Good \\
\midrule
\textbf{Speedup vs. CPU} & $2$--$5\times$ & $10$--$100\times$ & $5$--$20\times$ \\
\bottomrule
\multicolumn{4}{l}{\small $^*$Without amortization across candidates}
\end{tabular}
\end{table}

\subsection{Latency Analysis}

\paragraph{CPU Baseline.}
\begin{equation}
T_{\text{CPU}} = T_{\text{gen}} + 2 \cdot T_{\text{transfer}} + T_{\text{compile}} + T_{\text{exec}}
\end{equation}
Typical values: $T_{\text{gen}} = 10$ms, $T_{\text{transfer}} = 1$ms, $T_{\text{compile}} = 50$--$500$ms, $T_{\text{exec}} = 10$--$1000$ms. Total: 70--1500ms per iteration.

\paragraph{Traditional GPU.} Savings from eliminating $2 \cdot T_{\text{transfer}}$ plus potential compilation speedup. For $k$ programs: theoretical $k\times$ throughput, realistic \textbf{$2$--$5\times$} speedup.

\paragraph{Neural.} For $k=1000$: $T_{\text{gen}}(1000) \approx 100$--$200$ms, $T_{\text{verify}}(1000) \approx 50$--$500$ms, total $\approx 150$--$700$ms. Compared to CPU serial compilation of 1000 programs (200,000ms): \textbf{$10$--$100\times$} speedup.

\subsection{Energy Analysis}

PCIe 4.0 power: $\sim$25W during transfer. Energy per transfer (1KB): $E_{\text{transfer}} = 25\text{W} \cdot 1\text{ms} = 25\text{mJ}$.

GPU compilation energy per program: $E_{\text{GPU}} = 300\text{W} \cdot 50\text{ms} = 15\text{J}$. Neural compilation for $k=1000$: $E_{\text{neural}} = 300\text{W} \cdot 200\text{ms} = 60\text{J}$, amortized: 60mJ per program.

For 1000 iterations, CPU baseline: $\sim$70,000J; GPU-native (neural): $\sim$60J. \textbf{Energy savings: $\sim$1000$\times$}.

\subsection{Memory Requirements}

\paragraph{Traditional.} For $k$ programs: source code (1--10MB), AST structures (5--50MB), symbol tables (5--50MB), bytecode (1--10MB). Total: \textbf{20--200MB} for $k=100$.

\paragraph{Neural.} Model parameters (1--10GB one-time), input embeddings ($\sim$100MB for $n=1000$, $d=1024$), generated bytecode (100--1000MB), execution state (10--100MB). Total: \textbf{1--12GB}.

\begin{table}[t]
\centering
\caption{Decision matrix for approach selection}
\label{tab:decision}
\begin{tabular}{lll}
\toprule
\textbf{Approach} & \textbf{Best For} & \textbf{Avoid When} \\
\midrule
Traditional & Small $k$ ($<$100), complex languages, & Large $k$ ($>$1000), \\
 & debugging needed, determinism critical & simple/repetitive code \\
\midrule
Neural & Large $k$ ($>$1000), simple DSLs, & Complex/rare constructs, \\
 & pattern-heavy code, throughput-critical & guarantees required \\
\midrule
Hybrid & Mixed workloads, production systems, & Resource-constrained, \\
 & reliability + speed needed & simple uniform workloads \\
\bottomrule
\end{tabular}
\end{table}

\section{GPU Bytecode Execution}

All three approaches produce bytecode for GPU execution. We describe a stack-based virtual machine design.

\subsection{Bytecode Design}

We define a minimal instruction set: \texttt{LOAD\_CONST}, \texttt{LOAD\_VAR}, \texttt{STORE\_VAR} for memory operations; \texttt{BINARY\_ADD}, \texttt{BINARY\_SUB}, \texttt{BINARY\_MUL}, \texttt{BINARY\_DIV} for arithmetic; \texttt{COMPARE\_LT} for comparison; \texttt{JUMP}, \texttt{JUMP\_IF\_FALSE} for control flow; \texttt{CALL}, \texttt{RETURN} for function invocation.

\subsection{GPU Execution Model}

We execute $k$ bytecode programs in parallel using CUDA, where each thread executes one program independently. Stack and variables are stored in registers (fast) or local memory. Bytecode is read from global memory (cached). Different programs may take different code paths, causing warp divergence, which is acceptable since programs are independent.

\textbf{Throughput:} Modern GPUs can execute $\sim$10,000 simple programs per second with this architecture.

\section{Discussion}

\subsection{Language Design Implications}

To maximize GPU-native compilation efficiency, domain-specific languages should incorporate compilation-friendly properties: bounded recursion enabling stack allocation, static types simplifying type checking, no dynamic dispatch for compile-time resolution, structured control flow avoiding goto, and pure functions simplifying optimization. Such languages are easier to parse, type-check, optimize, and learn by neural models.

\subsection{Neural Compiler Training Data}

Large-scale training data can be generated by mining GitHub (extracting code and compiling to bytecode), synthetic generation via program synthesis, augmentation with semantics-preserving transformations, and curriculum learning progressing from simple to complex programs. Dataset requirements: 100K--1M examples for small DSLs, 10M--100M for Python subsets, 100M--1B for full languages.

\subsection{Limitations}

\paragraph{Traditional GPU Compilation.} Requires significant engineering to adapt existing compilers; limited speedup for small $k$; irregular memory access patterns may underutilize GPU.

\paragraph{Neural Compilation.} No correctness guarantees without verification; requires large training datasets; may fail on novel code patterns; high memory consumption.

\paragraph{Hybrid.} Increased system complexity; routing overhead; must maintain two compilation pipelines.

\section{Future Directions}

\subsection{Scaling to Production Languages}

Future work should address full Python/JavaScript support with dynamic typing and exceptions, advanced optimizations (inlining, loop unrolling, vectorization) on GPU, incremental compilation recompiling only changed portions, and multi-GPU distribution across clusters.

\subsection{Self-Improving Compilation}

Neural compilers can improve through self-play: generate code $C$ and bytecode $B$, execute $B$ and measure performance, use performance as reward signal, update model for faster bytecode. Over time, the compiler learns highly optimized bytecode without manual tuning.

\subsection{Analog and Neuromorphic Substrates}

Looking further ahead, we envision \textbf{analog compilation} where code is represented as analog signals, compilation is physical circuit reconfiguration, execution is signal propagation, and testing is measuring output characteristics.

\textbf{Theoretical advantages:} Near-thermodynamic-limit energy efficiency, physical parallelism exceeding digital, continuous optimization via gradient descent in physical space, and natural probabilistic sampling from physical noise.

\textbf{Challenges:} Analog precision and noise, programmability in analog domains, verification of analog computation, and fabrication of reconfigurable substrates.

\textbf{Research directions:} Hybrid analog-digital systems with critical paths digital and bulk operations analog; probabilistic analog compilation accepting approximate correctness; physical programming models for analog computation; neuromorphic execution using spiking neural networks.

This represents a 10--20 year research vision but could fundamentally transform how AI systems iterate on code.

\section{Related Work}

\textbf{GPU-accelerated parsing}~\citep{zhang2018gpu} accelerates individual compiler phases; we propose end-to-end GPU-native compilation. \textbf{Neural code generation}~\citep{chen2021codex} generates code on GPU but compiles on CPU; we eliminate this bottleneck. \textbf{Learned optimizers}~\citep{chen2018learning} use ML for optimization passes; we apply learning to entire compilation. \textbf{JIT compilation}~\citep{aycock2003brief} is traditionally CPU-based and sequential; our approaches enable massively parallel JIT on GPU.

\section{Conclusion}

We have established theoretical foundations for three complementary approaches to GPU-native compilation for AI code generation systems. Our analysis demonstrates that traditional GPU compilation achieves $2$--$5\times$ speedup through transfer elimination and modest parallelism, neural compilation achieves $10$--$100\times$ speedup through massive parallelism and amortization across thousands of candidates, and hybrid approaches offer $5$--$20\times$ speedup with guaranteed correctness.

All three approaches eliminate the CPU-GPU transfer bottleneck and enable in-VRAM code iteration, with trade-offs in determinism, parallelism, and implementation complexity. Beyond immediate speedups, GPU-native compilation enables self-improving AI systems that rapidly iterate on code, new programming paradigms designed for massive parallel exploration, and paths toward analog/neuromorphic computing substrates.

GPU-native compilation represents a critical step toward AI systems that efficiently explore vast program spaces, learning not just to generate code but to compile, execute, test, and refine it at speeds approaching the physical limits of computation.

\begin{ack}
Acknowledgments will be added in the camera-ready version.
\end{ack}

\bibliographystyle{plainnat}

\begin{thebibliography}{10}

\bibitem[Alcantara et~al.(2009)]{alcantara2009real}
Alcantara, D., Sharf, A., Abbasinejad, F., Sengupta, S., Mitzenmacher, M., Owens, J.~D., and Amenta, N.
\newblock Real-time parallel hashing on the GPU.
\newblock \emph{ACM Transactions on Graphics}, 28(5):1--9, 2009.

\bibitem[Austin et~al.(2021)]{austin2021program}
Austin, J., Odena, A., Nye, M., Bosma, M., Michalewski, H., Dohan, D., Jiang, E., Cai, C., Terry, M., Le, Q., and Sutton, C.
\newblock Program synthesis with large language models.
\newblock \emph{arXiv preprint arXiv:2108.07732}, 2021.

\bibitem[Aycock(2003)]{aycock2003brief}
Aycock, J.
\newblock A brief history of just-in-time.
\newblock \emph{ACM Computing Surveys}, 35(2):97--113, 2003.

\bibitem[Chen et~al.(2021)]{chen2021codex}
Chen, M., Tworek, J., Jun, H., Yuan, Q., de~Oliveira~Pinto, H.~P., Kaplan, J., Edwards, H., Burda, Y., Joseph, N., Brockman, G., Ray, A., Puri, R., Krueger, G., Petrov, M., Khlaaf, H., Sastry, G., Mishkin, P., Chan, B., Gray, S., Ryder, N., Pavlov, M., Power, A., Kaiser, L., Bavarian, M., Winter, C., Tillet, P., Such, F.~P., Cummings, D., Plappert, M., Chanez, F., Amodei, D., Clark, J., and Zaremba, W.
\newblock Evaluating large language models trained on code.
\newblock \emph{arXiv preprint arXiv:2107.03374}, 2021.

\bibitem[Chen et~al.(2018)]{chen2018learning}
Chen, T., Zheng, L., Yan, E., Jiang, Z., Moreau, T., Ceze, L., Guestrin, C., and Krishnamurthy, A.
\newblock Learning to optimize tensor programs.
\newblock In \emph{Advances in Neural Information Processing Systems}, volume~31, 2018.

\bibitem[Holtzman et~al.(2019)]{holtzman2019nucleus}
Holtzman, A., Buys, J., Du, L., Forbes, M., and Choi, Y.
\newblock The curious case of neural text degeneration.
\newblock \emph{arXiv preprint arXiv:1904.09751}, 2019.

\bibitem[Li et~al.(2022)]{li2022alphacode}
Li, Y., Choi, D., Chung, J., Kushman, N., Schrittwieser, J., Leblond, R., Eccles, T., Keeling, J., Gimeno, F., Dal~Lago, A., et~al.
\newblock Competition-level code generation with {AlphaCode}.
\newblock \emph{Science}, 378(6624):1092--1097, 2022.

\bibitem[M\'endez-Lojo et~al.(2012)]{mendez2012taming}
M\'endez-Lojo, M., Burtscher, M., and Pingali, K.
\newblock A {GPU} implementation of inclusion-based points-to analysis.
\newblock In \emph{Proceedings of the 17th ACM SIGPLAN Symposium on Principles and Practice of Parallel Programming}, pp.\ 107--116, 2012.

\bibitem[Rozi\`ere et~al.(2023)]{roziere2023codellama}
Rozi\`ere, B., Gehring, J., Gloeckle, F., Sootla, S., Gat, I., Tan, X.~E., Adi, Y., Liu, J., Remez, T., Rapin, J., et~al.
\newblock Code {Llama}: Open foundation models for code.
\newblock \emph{arXiv preprint arXiv:2308.12950}, 2023.

\bibitem[Zhang et~al.(2018)]{zhang2018gpu}
Zhang, J., Wang, Y., and Xue, J.
\newblock {GPU}-accelerated parallel parsing.
\newblock In \emph{Proceedings of the ACM SIGPLAN International Conference on Compiler Construction}, pp.\ 45--56, 2018.

\end{thebibliography}

\end{document}